# Identifying the origin of out-of-plane spin polarization in the noncollinear antiferromagnet Mn$_3$Ge

Mingxing Wu[1,2,*], Kouta Kondou[2], Taishi Chen[3], Satoru Nakatsuji[1,4,5], YoshiChika Otani[1,2,4,5,†]

[1]*The Institute for Solid State Physics, The University of Tokyo; Kashiwa, Chiba 277-8581, Japan*
[2]*Center for Emergent Matter Science, RIKEN; 2-1 Hirosawa, Wako 351-0198, Japan*
[3]*School of Physics, Southeast University; Nanjing 211189, China.*
[4]*Department of Physics, University of Tokyo; Hongo, Bunkyo-ku, Tokyo 113-0033, Japan*
[5]*Trans-scale Quantum Science Institute, University of Tokyo; Tokyo, Japan.*
(Dated: March 20, 2026)

The noncollinear antiferromagnets Mn$_3$Sn/Ge emerge as promising spin-current sources with both in-plane and out-of-plane spin polarizations, thereby enabling field-free magnetization switching. However, the microscopic origin of the out-of-plane spin polarization remains under debate, specifically whether it arises from the magnetic spin Hall effect (MSHE) or the spin swapping (SSW). Here, we comparatively evaluate the spin torques in single-crystal Mn$_3$Ge/Py bilayers with different crystallographic orientations using the ferromagnetic resonance technique. The distinct angular dependences of the measured spin-torque signals provide clear evidence for the bulk MSHE, which depends on antiferromagnetic order. In addition, we identify the antiferromagnetic-order independent component originating from the interfacial SSW. The coexisting MSHE and SSW, with comparable magnitudes, give rise to the out-of-plane spin polarization. Our study disentangles the origins of spin-torque generation in noncollinear antiferromagnets, providing valuable insights for their spintronic applications.

## I. INTRODUCTION

The spin Hall effect (SHE) enables charge-to-spin current conversion, and its resulting spin-orbit torque (SOT) offers an effective electrical means for manipulating magnetization [1–10]. In the conventional SHE framework, an in-plane charge current generates an out-of-plane spin current with an in-plane spin polarization dictated by symmetry. In a perpendicularly magnetized heterostructure, the accumulated spins exert SOTs on the adjacent magnetic layer, aligning the magnetization toward the film plane. However, deterministic magnetization reversal cannot be achieved because the torque vanishes when the magnetization becomes collinear with the spin polarization direction [11]. Consequently, an additional in-plane magnetic field is required to break the mirror symmetry and enable field-free deterministic magnetization switching.

Recent studies have demonstrated that an out-of-plane component of SHE can emerge in low-symmetry materials, generating an in-plane effective field that supports field-free magnetization switching. Such symmetry reduction may arise from the crystal structure, as in WTe$_2$ [12], MoTe$_2$ [13,14] and TaIrTe$_4$ [15,16], or from the magnetic structure, as in Mn$_3$Sn/Ge [17–21], Mn$_3$GaN [22] and Mn$_3$SnN [23]. Among these materials, the noncollinear antiferromagnets (AFMs) Mn$_3$Sn [24,25] and Mn$_3$Ge [26] stand out for the capabilities of field-free magnetization switching with low threshold current densities of approximately 5 MA/cm$^2$, manifesting their promise as efficient spin sources for low-power spintronic devices.

The microscopic origin of out-of-plane spin polarization in Mn$_3$Sn/Ge has been attributed to the magnetic SHE (MSHE) [18,27,28], which is odd under magnetization reversal. As illustrated in the upper panel of Fig. 1(a), when a charge current $J_c$ is applied, the conventional SHE generates an in-plane polarized spin current, while the MSHE simultaneously produces an additional spin accumulation with an out-of-plane polarization. Importantly, the MSHE is a bulk effect, and the sign reverses upon reversal of the magnetic octupole ordering [29]. In contrast, a recent study [30] argues that out-of-plane spin polarization in Mn$_3$Sn solely arises from spin swapping (SSW), i.e., the interchange between the flow direction and spin polarization of the spin current [31–33], which is insensitive to the magnetic octupole configuration [30]. As depicted in the lower panel of Fig. 1(a), a spin-polarized current in the ferromagnetic layer (Py) undergoes spin–orbit-coupled scattering in the adjacent material, resulting in spin swapping and thus giving rise to an out-of-plane spin polarization. Unlike the MSHE, the SSW depends on interfacial scattering and remains unchanged under magnetic octupole reversal [30]. Therefore, unambiguously identifying the origin of the out-of-plane spin polarization is crucial for understanding the mechanisms of spin–torque generation that enable field-free perpendicular magnetization switching using noncollinear AFMs [24,26,34].

In this work, we experimentally evaluate spin torques in the single-crystal Mn$_3$Ge/Py bilayer using the spin-torque ferromagnetic resonance (ST-FMR) technique. Our approach exploits the strong in-plane anisotropy of the kagome lattice, which constrains the magnetic octupole moments to rotate within the kagome plane. By performing comparative ST-FMR measurements on devices with different crystallographic orientations, we observed distinct angular dependences in the ST-FMR signals. These orientation-dependent features provide clear evidence for the MSHE, whose contribution depends on the orientation of the magnetic octupole. In addition, we also identify a magnetic-octupole-independent component, which we attribute to the SSW. Their experimental quantitative separation reveals that the coexisting MSHE and SSW have comparable magnitudes and together account for the observed out-of-plane spin polarization.

---

[*] mingxing.wu@mat.ethz.ch
[†] yotani@issp.u-tokyo.ac.jp

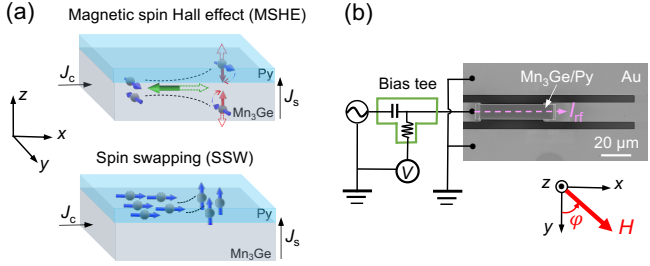

FIG 1. (a) Schematic illustrations of MSHE and SSW. The red arrows denote the out-of-plane spin polarization, whose sign reverses upon the reversal of magnetic octupole (green arrow). (b) Schematic of ST-FMR circuit with an image of the FIB fabricated device.

## II. METHODS

The Mn$_3$Ge single crystal strip was microfabricated using focused ion beam (FIB, FEI Scios) from a single crystal grown by the Bismuth flux method [35,36]. The strip with dimensions of 50 μm in length, 10 μm in width, and ~250 nm in thickness, was transferred to a coplanar waveguide which was patterned by Ti (5 nm)/Au (100 nm) on the SiO$_2$/Si substrate. We then cleaned the surface by Ar$^+$ milling and deposited 20 nm of Py using an electron-beam evaporator. The device was finally capped with 3 nm of AlO$_x$ to prevent oxidation. The measurements were performed using a GSG-type probe on a 3D vector magnetic field prober (Toei Scientific Industrial).

Figure 1(b) shows the scheme of the ST-FMR measurement with a SEM image of the fabricated device. An rf current (Keysight MXG N5183A) is applied in the device along x-direction in x-y plane. The spin current generated in the Mn$_3$Ge layer along the z-direction is injected into the adjacent Py layer and exerts spin torques. Finally, a dc voltage $V_{mix}$ is detected by a nanovoltmeter (Keithley 2182a), which originates from the anisotropic magnetoresistance (AMR) of Py. In general, the $V_{mix}$ consists of a symmetric component $V_S$ and an antisymmetric component $V_A$, given by [37]:

$$V_{mix} = V_S \frac{\Delta^2}{\Delta^2 + (H - H_r)^2} + V_A \frac{\Delta(H - H_r)}{\Delta^2 + (H - H_r)^2}, \quad (1)$$

where $\Delta$ is the linewidth, $H$ and $H_r$ are the applied and resonance magnetic fields. $V_S$ and $V_A$ are proportional to the magnitudes of the in-plane torque $\tau_{in}$ and out-of-plane torque

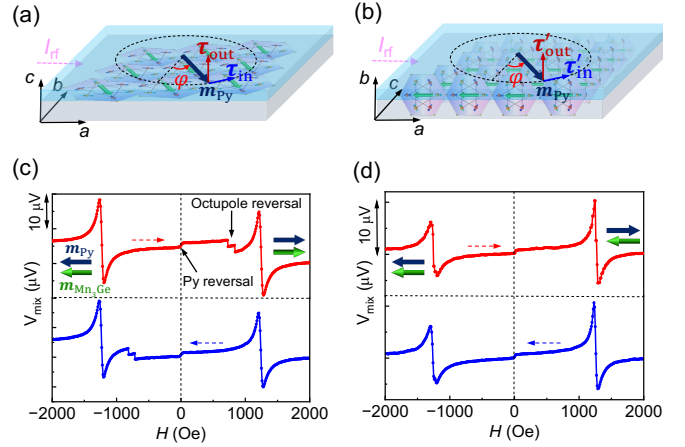

FIG 2. (a), (b) Setups for in-plane and out-of-plane devices. The magnetic octupole $m_{Mn_3Ge}$ follows $m_{Py}$ for the in-plane device shown in (a) but does not follow it for the out-of-plane device shown in (b) during the rotation of the applied magnetic field. $\tau_{in}^{(\prime)}$ and $\tau_{out}^{(\prime)}$ denote the current-induced in-plane and out-of-plane torques. (c), (d) Corresponding ST-FMR spectra under $\varphi = 45°$. The frequency and power of input rf current are 11 GHz and 10 mW.

$\tau_{out}$, respectively [37]. They are expressed respectively as [30,38]:

$$V_S = -\frac{I_{rf}}{2}\left(\frac{dR}{d\varphi}\right)\frac{1}{\alpha(2\mu_0 H_r + \mu_0 M_{eff})}\tau_{in}, \quad (2)$$

$$V_A = -\frac{I_{rf}}{2}\left(\frac{dR}{d\varphi}\right)\frac{\sqrt{1 + \frac{M_{eff}}{H_r}}}{\alpha(2\mu_0 H_r + \mu_0 M_{eff})}\tau_{out}, \quad (3)$$

where $I_{rf}$ denotes the input rf current, $R$ is the sample resistance, $\alpha$ is the Gilbert damping coefficient, and $M_{eff}$ is the effective magnetization of Py.

## III. RESULTS AND DISCUSSIONS

We microfabricated an in-plane device with the crystallographic c-axis oriented along the z-direction [Fig. 2(a)] and an out-of-plane device with c-axis along y-direction [Fig. 2(b)]. In both configurations, the rf current is applied along a-axis. The injected rf current generates a spin current that exerts an in-plane torque ($\tau_{in}^{(\prime)}$) and an out-of-plane torque ($\tau_{out}^{(\prime)}$) on the Py layer. The representative ST-FMR spectra measured under $\varphi = 45°$ are shown in Figs. 2(c) and (d). We measured hysteresis by sweeping the magnetic field from negative to positive and backward. The magnetization of Py layer is readily reversed under a small magnetic field which corresponds to the voltage change around zero field. In contrast, the reversal behavior of the magnetic octupoles differs between the in-plane and out-of-plane devices. Sharp voltage drops are observed at $H = \pm 720$ Oe for the in-plane device, which are ascribed to the reversal of magnetic octupole in Mn$_3$Ge, as shown in Fig. 2(c) [39]. However, no such reversal occurs in the out-of-plane device in Fig. 2(d). It indicates that magnetic octupole moments are pinned and irreversible within the applied magnetic field range. This distinct magnetic octupole reversal makes it possible to verify the MSHE. Specifically, the magnetic octupole and the magnetization of Py remain parallel in the vicinity of the resonance field, resulting in the identical $V_{mix}$ signals under negative and positive field regimes, as shown in Fig. 2(c). In contrast, in Fig. 2(d), the relative orientation between magnetic octupole and Py magnetization changes from parallel to antiparallel as the magnetic field is swept from negative to positive. Correspondingly, the $V_{mix}$ spectra differ for opposite field polarities. These comparative measurements suggest that the spin-torque response depends sensitively on the magnetic octupole order, implying the existence of MSHE.

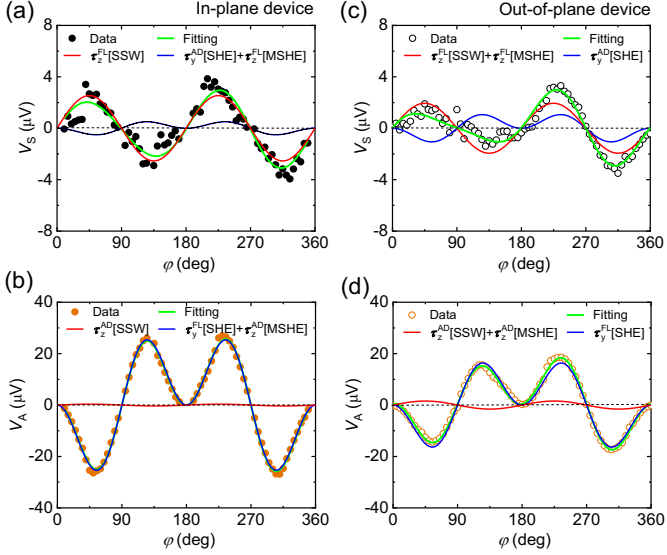

FIG. 3. $V_S$ and $V_A$ extracted from the angular dependence of ST-FMR measurements for the in-plane devices shown in (a) and (b), and for out-of-plane devices shown in (c), (d). The fittings are carried out using Eqs. (6) and (7) for the in-plane devices and using Eqs. (8) and (9) for the out-of-plane devices.

The spin torques can originate from the conventional SHE with in-plane spin polarization arising from SHE, as well as from the MSHE and SSW with out-of-plane spin polarizations. Each contribution can be further decomposed into an anti-damping-like torque $\tau^{AD}$ and a field-like torque $\tau^{FL}$. As a result, $\tau_{in}$ and $\tau_{out}$ can be rewritten as:

$$\tau_{in} = \tau_y^{AD}[SHE] + \tau_z^{FL}[MSHE] + \tau_z^{FL}[SSW], \quad (4)$$

$$\tau_{out} = \tau_y^{FL}[SHE] + \tau_z^{AD}[MSHE] + \tau_z^{AD}[SSW], \quad (5)$$

where the superscript denotes the type of spin torque, the subscript denotes the spin polarization direction, and [...] depicts the origin of spin torque. Here, we do not include the spin torque induced by spin current arising from the $x$ spin polarization [30,38], as its contribution is found to be negligibly small in our fittings.

To identify each spin-torque contribution, the anti-damping-like and field-like torques are further expressed as $\tau_i^{AD}[...] \propto \mathbf{m} \times (\mathbf{m} \times \mathbf{S}_i)$ and $\tau_i^{FL}[...] \propto \mathbf{m} \times \mathbf{S}_i$, respectively, where $\mathbf{S}_i$ is the spin polarization with $i = y, z$, and $\mathbf{m}$ denotes the magnetization of Py. We now derive the analytical expression for each contribution. First, the conventional SHE is independent of the magnetic octupole of Mn$_3$Ge. Accordingly, $\tau_y^{AD}[SHE]$ and $\tau_y^{FL}[SHE]$ exhibit sinusoidal functions with respect to $\varphi$ in both in-plane and out-of-plane devices, since the Py magnetization always follows the applied magnetic field. Second, the MSHE depends explicitly on the magnetic octupole, and can be approximately described by the cross product of the magnetic octupole order and $y$-direction [18,27,28,30]. As a result, both $\tau_z^{FL}[MSHE]$ and $\tau_z^{AD}[MSHE]$ scale as $\sin\varphi$ in the in-plane device where the magnetic octupole aligns to the magnetic field, whereas they remain constant for the out-of-plane device due to the pinning of the magnetic octupole. Third, the SSW originates from interfacial scattering and is insensitive to the magnetic octupole of Mn$_3$Ge. The associated torques remain unchanged upon the rotation of the magnetic field. Thus, each torque component can be identified by its symmetry with respect to the angle $\varphi$ and the magnetic octupole reversal. Meanwhile, under our measurement geometry, the AMR of Py gives $\frac{dR}{d\varphi} \propto sin2\varphi$. Based on these analyses, the final expression of $V_S$ and $V_A$ can be formulated as follows.

For an in-plane device:

$$V_S(\varphi) \propto -\sin2\varphi\left(\tau_y^{AD}[SHE]\sin\varphi + \tau_z^{FL}[MSHE]\sin\varphi + \tau_z^{FL}[SSW]\right), \quad (6)$$

$$V_A(\varphi) \propto -\sin2\varphi\left(\tau_y^{FL}[SHE]\sin\varphi + \tau_z^{AD}[MSHE]\sin\varphi + \tau_z^{AD}[SSW]\right). \quad (7)$$

For an out-of-plane device:

$$V_S(\varphi) \propto -\sin2\varphi(\tau_y^{AD}[SHE]\sin\varphi + \tau_z^{FL}[MSHE] + \tau_z^{FL}[SSW]), \quad (8)$$

$$V_A(\varphi) \propto -\sin2\varphi(\tau_y^{FL}[SHE]\sin\varphi + \tau_z^{AD}[MSHE] + \tau_z^{AD}[SSW]). \quad (9)$$

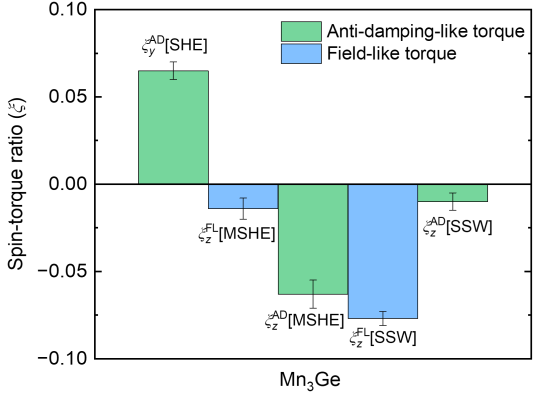

FIG. 4 Spin torque ratios for SHE, MSHE and SSW extracted from the fits of angular-dependent ST-FMR measurements. The error bars denote the standard deviation from the fitting.

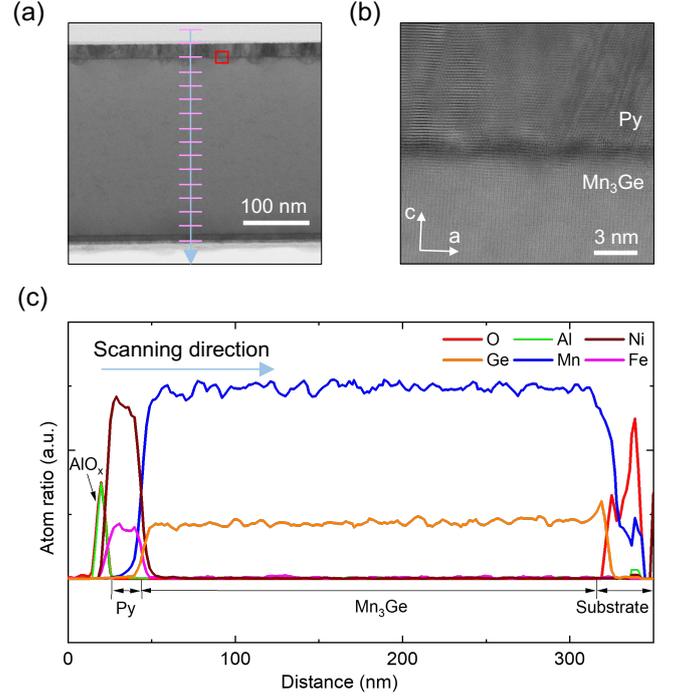

FIG. 5 (a) Cross-sectional bright-field STEM image of an in-plane device. The arrow depicts the scanning direction of the EDX mapping. The red rectangle shows the region for the high-resolution observation. (b) High-resolution STEM image for the interface of $Mn_3Ge$ and Py layers. (c) Element distribution along the thickness direction from EDX mapping. There is no observable compositional gradient in our device.

Following the framework discussed above, we measured the angular dependence of ST-FMR for both in-plane and out-of-plane devices as shown in Fig. 3(a-d). The magnetic field was initially set to −2500 Oe and subsequently scanned from −500 to −2000 Oe at different $\varphi$ angles. The $V_S$ and $V_A$ were fitted using Eqs. (6) and (7) for the in-plane devices and using Eqs. (8) and (9) for the out-of-plane devices. The fitting results clearly indicate that all three mechanisms, namely the SHE, MSHE, and SSW, coexist and contribute to the observed spin torques. For in-plane spin torques, $\tau_y^{AD}[SHE]$ extracted from Fig. 3(c) is sizeable while $\tau_y^{AD}[SHE] + \tau_z^{FL}[MSHE]$ in Fig. 3(a) is reduced. This behavior suggests that $\tau_z^{FL}[MSHE]$ partially compensates $\tau_y^{AD}[SHE]$, in agreement with the previous report for $Mn_3Sn$ [28]. In addition, $\tau_z^{FL}[SSW]$ also exhibits an opposite sign relative to $\tau_y^{AD}[SHE]$. As a result, the magnitude of the in-plane torque originating from the SHE is suppressed by the competing contributions from both MSHE and SSW.

To quantitatively compare the individual spin-torque contributions, we define the spin-torque ratio as $\xi_i^{AD(FL)}[\ldots] = \frac{\tau_i^{AD(FL)}[\ldots]}{\tau_y^{FL}[SHE]}$, where $\tau_y^{FL}[SHE]$ is taken as the reference torque because it predominantly originates from the current-induced Oersted field. For instance, $\xi_y^{AD}[SHE] = \frac{\tau_y^{AD}[SHE]}{\tau_y^{FL}[SHE]}$ can be directly extracted from the fits to Figs. 3(c) and 3(d), yielding a value of $0.065 \pm 0.005$ for the $Mn_3Ge$ single crystal. Similarly, $\xi_z^{FL}[SSW]$ and $\xi_z^{AD}[SSW]$ are determined to be $-0.077 \pm 0.004$ and $-0.01 \pm 0.05$ from the fits of Figs. 3(a) and 3(b), respectively, where we approximate $\tau_y^{FL}[SHE] \approx \tau_y^{FL}[SHE] + \tau_z^{AD}[MSHE]$ as $\tau_z^{AD}[MSHE]$ is only about 2% of $\tau_y^{FL}[SHE]$ [28]. Furthermore, from the fits of Figs. 3(c) and 3(d), we extract $\xi_z^{FL}[MSHE] + \xi_z^{FL}[SSW] = -0.091 \pm 0.004$ and $\xi_z^{AD}[MSHE] + \xi_z^{AD}[SSW] = -0.073 \pm 0.006$. Consequently, one obtains $\xi_z^{FL}[MSHE] = -0.014 \pm 0.006$ and $\xi_z^{AD}[MSHE] = -0.063 \pm 0.008$, by assuming the same contribution of SSW for both in-plane and out-of-plane devices due to interface similarity from the same fabrication process [30,40–43].

The extracted spin-torque ratios are summarized in Fig. 4. The spin torques arising from the MSHE are comparable in magnitude but have opposite signs to those arising from the conventional SHE. Notably, previous measurements on the out-of-plane spin polarization in $Mn_3Sn$ [28,30,38] and $Mn_3Ge$ [26] were performed on devices only with kagome in-plane crystallographic orientation. In this configuration, the MSHE and SHE cannot be distinguished by symmetry analysis as they exhibit the same angular dependences, as shown in Eqs. (6) and (7). Alternatively, microscopic calculations based on quantum kinetic theory were employed to evaluate the MSHE contribution [28]. In contrast, we can experimentally determine the magnitude of the MSHE by comparatively measuring devices with in-plane and out-of-plane crystallographic orientations.

In addition to the MSHE, the sizeable magnetic-octupole-independent SSW is also observed, which generates a giant in-plane field-like torque. The SSW arises from interfacial scattering, as demonstrated in various nonmagnetic-ferromagnetic heterostructures [40–43]. The relative strength of the in-plane spin torques arising from SHE and SSW is determined by the competing spin-orbit coupling and interfacial disorder [32,33]. Specifically, SSW is pronounced only in systems with weak spin-orbit coupling and low interfacial disorder. $Mn_3Ge$ is known to have weak spin-orbit coupling [44]. We now examine the interfacial quality of the Mn$_3$Ge/Py bilayer. Figures 5(a) and (b) show the cross-sectional bright-field STEM images for an in-plane device. The Mn$_3$Ge layer exhibits high crystallinity with the *c*-axis oriented normal to the surface. Moreover, a crystallized interface with a thickness of a few atomic layers is observable between the Mn$_3$Ge and Py layers. This high-quality interface is consistent with and likely responsible for the pronounced SSW in the Mn$_3$Ge/Py bilayer. Besides the SSW, symmetry breaking arising from compositional gradient [45–47] can also induce an out-of-plane spin polarization that is insensitive to the magnetic structure. To examine this possibility, we performed STEM-EDX measurements to assess the presence of a compositional gradient in Mn$_3$Ge that could have been introduced during the FIB microfabrication process. We mapped the element distribution along the thickness direction and plotted the atom ratio, as shown in Fig. 5(c). The elements are homogeneously distributed in the regions from the AlO$_x$ capping layer, Py, Mn$_3$Ge, to the substrate. We do not observe any composition gradient, and this mechanism can thus be ruled out. Consequently, we identify the SSW as the dominant origin of the magnetic-octupole-independent out-of-plane spin polarization.

## IV. CONCLUSION

We carried out ST-FMR measurements on single-crystal Mn$_3$Ge/Py bilayers with different crystallographic orientations to elucidate the origins of out-of-plane spin polarization in the noncollinear AFM Mn$_3$Ge. In the in-plane geometry, the magnetic octupole follows the applied magnetic field, whereas in the out-of-plane geometry, it is effectively pinned. This provides a controlled way to isolate the magnetic-octupole-dependent MSHE from the magnetic-octupole-independent SSW contribution. The angular dependence of ST-FMR analysis reveals that the two mechanisms coexist and jointly account for the out-of-plane spin polarization. Our study provides important insights into spin-torque mechanisms responsible for current-induced magnetization switching [24,26,34,48,49] and domain-wall motion [50,51] based on noncollinear AFMs.

## ACKNOWLEDGMENT

This work is financially supported by JST-CREST (no. JPMJCR18T3), JST-Mirai Program (no. JPMJMI20A1), Japan Science and Technology Agency. T. C is supported by the National Key Research and Development Program of China (2023YFA1406600), the National Natural Science Foundation of China (no. 12274068). M. W. would like to acknowledge support from JSPS "Research Program for Young Scientists (no. 21J21461).